\documentclass[%
reprint,
superscriptaddress,
 amsmath,amssymb,
 aps,
 prl,
]{revtex4-2}

\usepackage{graphicx}% Include figure files
\usepackage{dcolumn}% Align table columns on decimal point
\usepackage{bm}% bold math
\begin{document}

\preprint{APS/123-QED}

\title{
Topological Linking Drives Anomalous Thickening of Ring Polymers In Weak Extensional Flows  
}

\author{Thomas C. O'Connor}
\email{toconno@sandia.gov}
 \affiliation{Sandia National Laboratories, Albuquerque, New Mexico, 87185, USA}

\author{Ting Ge}%
% \email{Second.Author@institution.edu}
\author{Michael Rubinstein}
\affiliation{Department of Mechanical Engineering and Materials Science, Duke University, Durham, North Carolina, 27708, USA}

\author{Gary S. Grest}
\affiliation{Sandia National Laboratories, Albuquerque, New Mexico, 87185, USA}

\date{\today}% It is always \today, today,
             %  but any date may be explicitly specified
\begin{abstract}
Molecular dynamics simulations confirm recent extensional flow experiments showing ring polymer melts exhibit strong extension-rate thickening of the viscosity at Weissenberg numbers $Wi<<1$. Thickening coincides with the extreme elongation of a minority population of rings that grows with $Wi$. The large susceptibility of some rings to extend is due to a flow-driven formation of topological links that connect multiple rings into supramolecular chains.
Links form spontaneously with a longer delay at lower $Wi$ and are pulled tight and stabilized by the flow.
Once linked, these composite objects experience larger drag forces than individual rings, driving their strong elongation. The fraction of linked rings depends non-monotonically on $Wi$, increasing to a maximum when $Wi\sim1$ before rapidly decreasing when the strain rate approaches $1/\tau_e$.
\end{abstract}

%\keywords{Suggested keywords}%Use showkeys class option if keyword
                              %display desired
\maketitle
Non-concatenated ring polymers do not entangle like linear polymers, and they tend to have lower Newtonian viscosities than linear melts of similar molecular weight \cite{Rubinstein1986,Obukhov1994,Kapnistos2008, halverson11a,halverson11b,Ge2016}. 
However, recent experiments by Huang et al. \cite{Huang2019} show that ring melts are extraordinarily sensitive to extensional flow, exhibiting large, nonlinear growth in the extensional viscosity $\eta_{ex}$ even when the strain rate $\dot{\varepsilon}$ is small compared to the reciprocal ring relaxation time $1/\tau_{ring}$ in equilibrium.
Such trends are usually due to polymer conformations elongating in the flow field \cite{halverson11b,halverson12,OConnor2018PRL,Mortensen2018,Zhou2019}, but analytic and numerical models for linear response predict this shouldn't happen when $\dot{\varepsilon}<<1/\tau_{ring}$ \cite{Ge2016,halverson11b}.
Thus, it appears that new physics emerges and changes how rings dissipate energy in uniaxial extensional flows. 

Huang et al. observed an apparent correspondence between the high-rate extensional viscosity $\eta_{ex}$ of ring and linear melts of Polystyrene (PS) with the same $M_w=185$ kg/mol, leading them to propose that rings transition to behaving like linear chains as they elongate \cite{Huang2019}. 
However, prior simulation studies of unentangled ring and linear melts under planar extension did not observe a clear correspondence at large $Wi$ \cite{Yoon2016}.
In addition, scaling arguments and simulations of linear melts predict that $\eta_{ex}$ should scale as $\sim N^2$ for extended linear chains \cite{OConnor2018PRL} and as $\sim N^2/4$ for rings if they behave like linear chains. 
Thus, extended rings of $N$ monomers should behave like linear chains with $N/2$ monomers, but this was not observed in experiments.

\begin{figure}
    \centering
    \includegraphics[width=\linewidth]{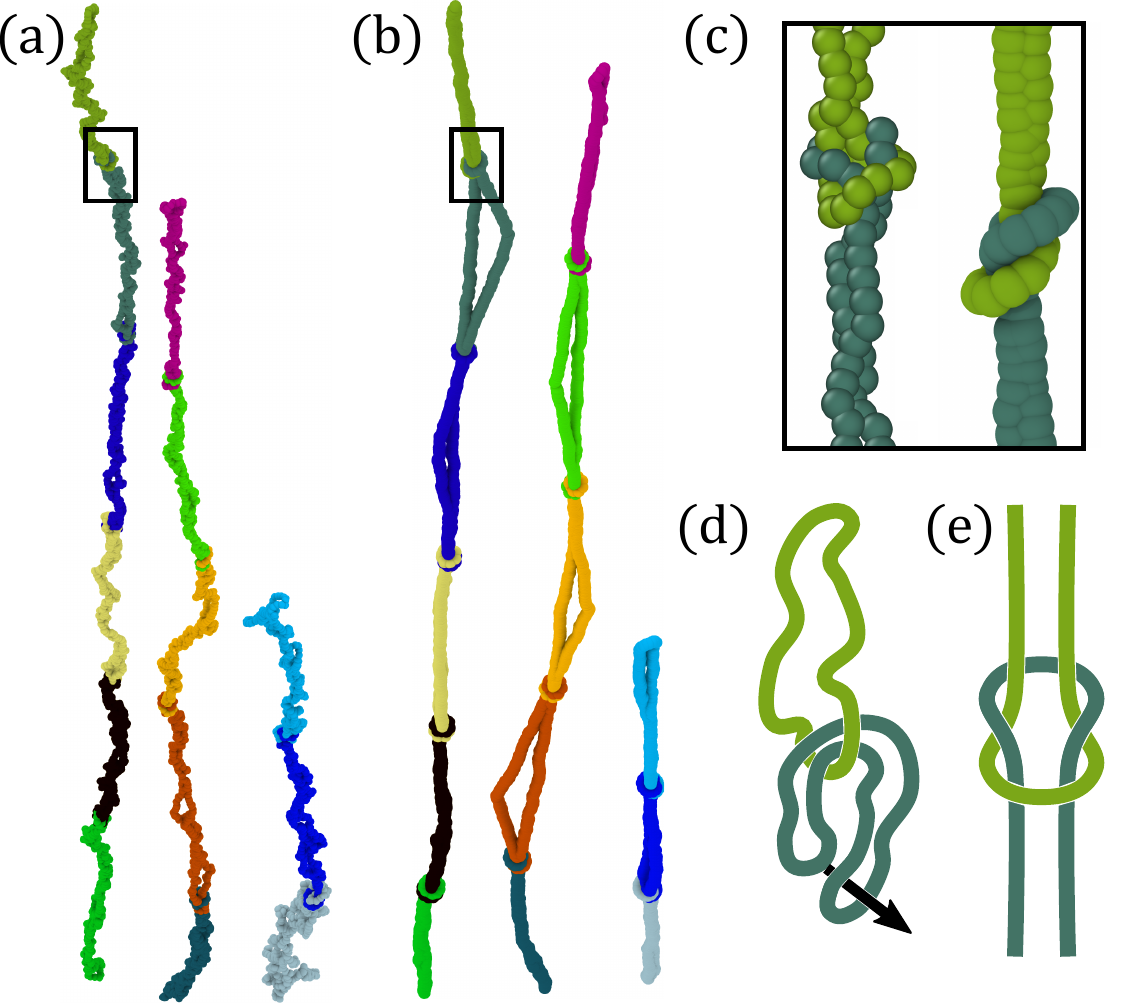}
    \caption{(a) Elongated $N=400$ rings from the terminal state of the $Wi=0.16$ flow. 
    (b) Chains from (a) after PPA, revealing tight-links connecting rings into supramolecules. 
    (c) Close-up of a tight-link framed in (a) and (b) before and after PPA.
    (d) Diagram of the formation of a link and (e) the final structure once it is pulled taut.
    \label{fig:configs}
    }
\end{figure}

Here, we use MD simulations to show that the unintuitive rheology of rings under uniaxial extension is driven by a new topological linking mechanism that assembles rings into supramolecular daisy-chains during flow (Fig. \ref{fig:configs}).
We simulate start-up uniaxial extensional flows for ring and linear polymer melts at several molecular weights.
Simulations reproduce the strong sensitivity of rings to uniaxial extension reported by Huang et al., and also reveal qualitative differences between the elongation dynamics of rings and linear chains.
While chains in linear melts elongate independently during extension, the susceptibility of rings to assemble into supramolecules can drive a minority of rings to elongate and dominate the nonlinear viscosity, even when $Wi<<1$.
While many quantum materials in applied magnetic fields develop topological interactions that drive nonlinear response, ring polymers linking during extension is the first instance of a similar effect in a classical polymer liquid. 

We model polymers with a standard semi-flexible bead-spring model that has been used to study ring melts in equilibrium and under shear \cite{Kremer1990,halverson12}.
All beads interact with a truncated Lennard-Jones (LJ) potential and results are presented in reduced LJ units.
Rings of $N=200$, 400, and 800 beads are bound together with a FENE potential with mean bond length $b \approx 0.96$.
The chain stiffness is controlled by a bond bending potential $k_{\theta} (1- \cos \theta)$, where $\theta$ is the angle between successive bonds.
In this study, we fix $k_{\theta}=1.5$, which gives a Kuhn segment length $n_K\sim2.88$ beads and an entanglement strand containing $N_e\approx28$ beads \cite{Moreira2015}.

$M$ ring or linear polymers of length $N$ are equilibrated at a temperature $T=1$ and density $\rho=0.85$ using standard methods \cite{Auhl2003}.
The majority of our analysis considers a ring melt with $M=5000$ and $N=400$ ($Z=N/N_e\approx14$) that most closely correspond to the experimental results of Huang et al. on polystyrene (PS) of $M_w=185$ kg/mol \cite{Huang2019}.
In order to compare trends for different molecular weights and compare rings to linear melts, we also model ring melts with: $M=200$, $N=200$; $M=5000$, $N=800$; and linear melts with $M=1640$, $N=112$; $M=250$, $N=200$; $M=400$, $N=400$.
All simulations are performed using LAMMPS with a time-step $\Delta t=0.007$.

Melts are elongated at constant Hencky strain rate $\dot\varepsilon\equiv\partial\ln\Lambda/\partial t$ with $\Lambda$ the stretch along the z-axis. 
Upper limits of $\dot\varepsilon$ are chosen to avoid covalent bond stretching. 
Since polymers are nearly incompressible, the two perpendicular directions contract uniformly to preserve volume.
Generalized Kraynik-Reinelt boundary conditions are used to prevent the simulation box from collapsing in the contracting directions \cite{Dobson2014, Nicholson2016, OConnor2018PRL, OConnor2019}.
Flow is maintained by integrating the SLLOD equations of motion with a thermostat damping time of $10$ \cite{evans84,daivis06}.
Like experiments, we record the start-up extensional viscosity $\eta_{ex}(t)=\sigma_{ex}/\dot{\varepsilon}$ during flow, where the extensional stress $\sigma_{ex}=\sigma_{zz}-\sigma_{rr}$ is the difference in stress between the extension axis (z-axis) and the transverse directions.

Experiments typically analyze data from different chemistires by comparing them in reduced units derived from linear response.
The same approach can be used to compare bead-spring simulations to the experiments of Huang et al. \cite{Huang2019}
For bead-spring simulations, $\tau_{ring}(N)$ is taken from Ge and Rubinstein's fits of simulation data to the Fractal-Loopy-Globule (FLG) model\cite{Ge2016}.
For the experiments, we use the empirically measured $\tau_{ring}=52.6$ s.
Strain rates are reported as a dimensionless Weissenberg number $Wi=\dot{\varepsilon}\tau$, with $\tau=\tau_{ring}$ for rings and $\tau=\tau_R=\tau_{e}Z^2$ for linear melts.
$\tau_R$ is the Rouse time which is the longest dissipative relaxation time of a linear chain in the absence of entanglements
It can be related to the entanglement time $\tau_e$ and the number of entanglements per chain $Z$. Prior studies on this bead-spring model have measured $\tau_{e}\approx1.98\times10^3$ \cite{Moreira2015}. 

\begin{figure}
    \centering
    \includegraphics[width=\linewidth]{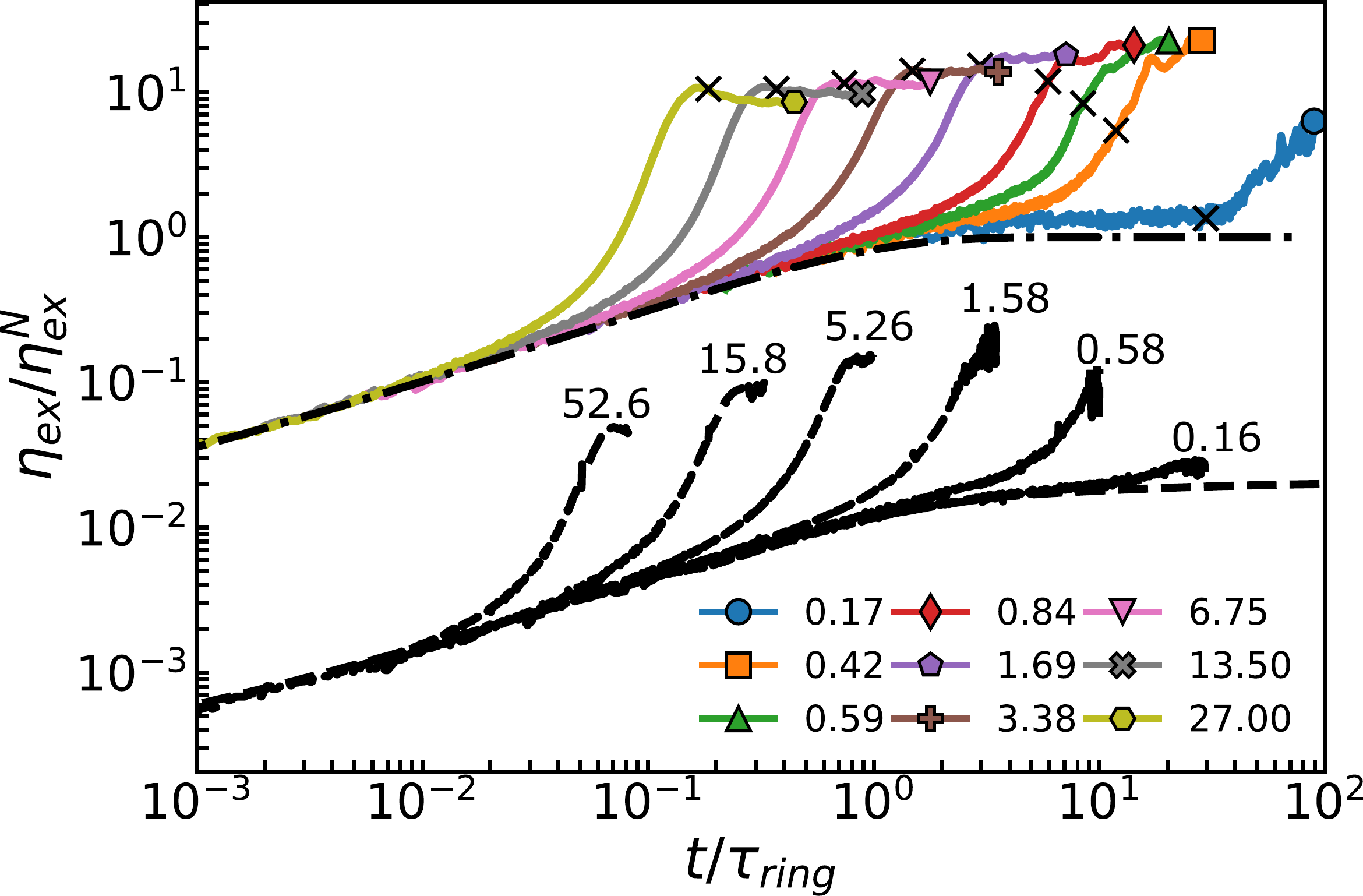}
    \caption{Start-up extensional viscosity $\eta_{ex}/\eta_{ex}^N$ versus $t/\tau_{ring}$ for a N=400 ring melt (upper data) and PS experiments of Huang et al. \cite{Huang2019} (lower data) at the indicated $Wi$. Dashed and dot-dashed lines show linear envelopes for experiments and simulations, respectively.
    Both simulations and experiments are for $Z\approx14$ and experiments are shifted down by a factor of 100 for clarity.
    Experiments terminate at lower Hencky strains ($\varepsilon \sim5$) then simulations ($>12$).
    Crosses on simulated curves indicate $\varepsilon=5$.
    \label{fig:startup}
    }
\end{figure}

Figure \ref{fig:startup} plots the normalized extensional viscosity $\eta_{ex}/\eta_{ex}^N$ versus reduced time $t/\tau_{ring}$ for a bead-spring melt of rings alongside the experimental curves from Huang et al. \cite{Huang2019}.
Both simulation and experiment use rings with $Z\sim14$.
$\eta_{ex}^N$ is the Newtonian viscosity taken from the plateau of the linear viscoelastic envelope (LVE).
The experimental LVE is take from Ref. \citenum{Huang2019}, while the simulated LVE is taken from Eq. 49 of Ref. \citenum{Ge2016}.

The simulations reproduce the qualitative trends with $Wi$ that are seen in experiments.
Both show a large nonlinear growth in $\eta_{ex}$ above the Newtonian viscosity even for $Wi<<1$.
Extraordinarily, even for the lowest $Wi=0.16$, thickening above $\eta^N_{ex}$ is observed in simulations after $\sim 30\tau_{ring}$.
As there is no uncertainty about the homogeneity of composition or flow in the simulated melts, our data support Huang et al.'s conclusion that ring melts exhibit an anomalously large (and seemingly delayed) sensitivity to extensional flow.

Comparing the experimental curves to simulations in Figure \ref{fig:startup} shows an apparent difference in the trend of terminal viscosity with $Wi$. 
Simulations resolve steady-state plateaus in $\eta_{ex}$ for $Wi>0.16$ that are a factor $\sim10$ larger than $\eta_{ex}^N$ and that weakly decrease with increasing $Wi$.
In contrast, experiments, particularly at lower $Wi$, do not resolve clear plateaus and give terminal viscosities that decrease towards $\eta_{ex}^N$ as $Wi$ decreases.
These differences appear to be due to different maximum strains in simulations and experiments. 
The average maximum strain in the experiments is $\varepsilon\sim5$, while our ring simulations all reach $\varepsilon>12$.
The simulation curves are marked with an ``X'' at the points corresponding to $\varepsilon=5$.
Simulations truncated at $\varepsilon=5$ show the same qualitative trends with $Wi$ as seen in experiments.
This suggests that the ring experiments at low $Wi$ are likely not resolving steady-state viscosities, so comparing these values to steady-state $\eta_{ex}$ for linear melts is difficult.
Further, due to the apparent instability of rings towards thickening at times $t>>\tau_{ring}$, current experiments may not be able to reach sufficiently large strains to resolve steady-states at low $Wi$.

\begin{figure}
    \centering
    \includegraphics[width=\linewidth]{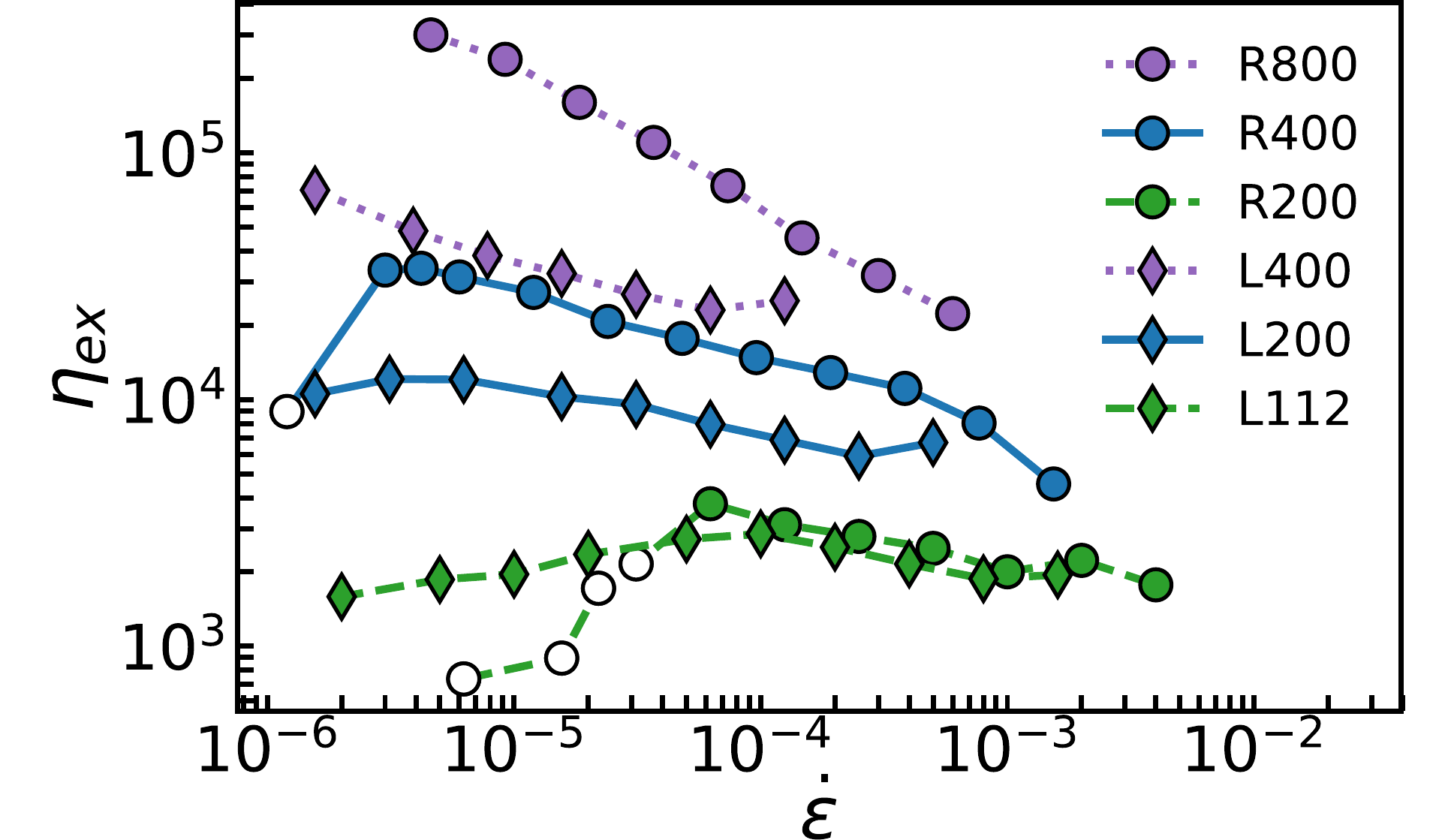}
    \caption{Steady-state $\eta_{ex}$ versus $\dot{\varepsilon}$ for all three ring molecular weights compared to linear melt data with $N_\ell \approx N/2$. $N_\ell$ are chosen so elongated rings and linears have similar dimensions at full extension.
    Unfilled symbols are points that did not reach steady-state by $\varepsilon=12$.
    \label{fig:steady}
    }
\end{figure}

Figure \ref{fig:steady} compares steady-state $\eta_{ex}$ versus $\dot{\varepsilon}$ for three ring and three linear molecular weights. 
For the smallest $N=200$ rings and $N=112$ linears, the steady $\eta_{ex}$ converge once $\dot{\varepsilon}>10^{-4}$.
However, this correspondence disappears with increasing molecular weight $N$.
$N=400$ rings and $N=200$ linears show a similar decrease in $\eta_{ex}$ with increasing $\dot{\varepsilon}$, but ring $\eta_{ex}$ are consistently $\sim2-3$ times larger until $\dot{\varepsilon}>2\times10^{-4}$.
At higher rates, the linear melt stops thinning while $\eta_{ex}$ for rings continue to decrease. 
$N=800$ rings and $N=400$ linears exhibit similar behavior, however, the separation between ring and linear $\eta_{ex}$ at intermediate rates has grown to a factor $\sim4-5$.

Prior studies found $\eta_{ex}\sim N^2$ for linear chains at large $Wi$ \cite{OConnor2018PRL}, the rings do not follow this scaling over our range of $\dot{\varepsilon}$.
Thus, the viscosity of an elongated melt of $N$-rings is not simply related to that of an elongated linear melt.
The ring-linear correspondence observed by Huang et al. for PS with $Z\sim14$ may be a crossing of two different $N$ dependencies at large $Wi$, like our $200$-ring and $112$-linear data appear to be.

The different trends in $\eta_{ex}$ in Figure \ref{fig:steady} imply that rings and linear chains are governed by different physics in extensional flows.
This becomes clear when we examine the conformations of chains during flow.
A common measure of linear chain elongation is the rms-magnitude of the end-end vector $\vec{R}$ connecting the two chain ends.
Rings do not have chain ends, so instead we define an effective end-end distance for a ring as the maximum distance spanned by any pair of monomers on the ring that are separated by $N/2$ bonds.
With this defintion, an $N$-ring at full extension has approximately the same $R$ as an $(N/2)$-linear chain.

\begin{figure}
    \centering
    \includegraphics[width=\linewidth]{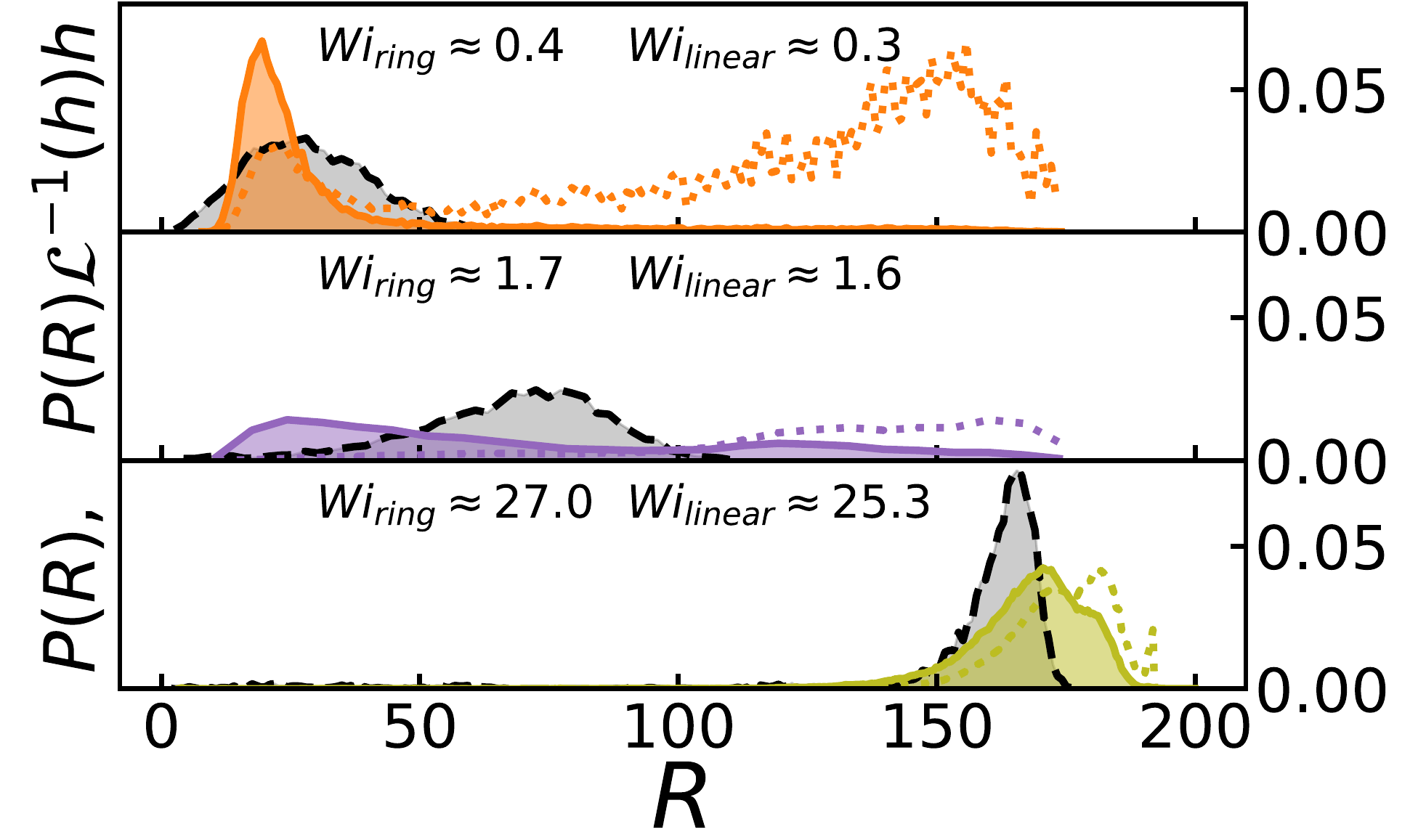}
    \caption{Steady-state $P(R)$ for $N=400$ rings (solid) and $N=200$ linears (dashed) at three $Wi$.
    Relative stress contributions $P(R)\mathcal{L}^{-1}(h)h$ for rings (dotted) are scaled to fit on the same axes. 
    \label{fig:rdist}
    }
\end{figure}

Figure \ref{fig:rdist} shows steady-state span distribution functions $P(R)$ at three values of $Wi$ for $N=400$ rings (solid lines) and $N=200$ linear chains (dashed lines).
The $Wi$ for the ring and linear chains are approximately the same for the three cases.
For linear melts, $P(R)$ exhibits a narrow distribution that shifts to larger $R$ with increasing $Wi$. 
This is consistent with the assumption that a deformed entanglement network transmits macroscopic strains independently to each chain.
However, the distributions for rings exhibit a fundamentally different behavior.
At the low $Wi=0.3$, $P(R)$ for the rings develop a long tail corresponding to a small number of highly stretched rings.
The tail grows with increasing $Wi$, producing a broad distribution for $Wi\sim1$.
For $Wi>>1$, the flow becomes fast enough to elongate all rings and $P(R)$ narrows to a single peak centered at large $R$. 

At low $Wi$ the fraction of highly elongated rings is small, but their contribution to the extensional stress is substantial.
This is due to the nonlinear increase in the entropic tension of a ring with increasing $R$. 
If we approximate the elongated rings to be under uniform tension, we can estimate the nonlinear increase in entropic tension as $\propto \mathcal{L}^{-1}(h)h$, where $\mathcal{L}^{-1}$ is the inverse Langevin function and $h=R/(bN/2)$ is the extension ratio of the ring.
In Figure \ref{fig:rdist}, the relative contribution of the rings to the stress $\propto P(R)\mathcal{L}^{-1}(h)h$ is plotted with a dotted line. 
For $Wi=27$, the distributions for rings and linears are similar and all rings contribute substantially to the stress.
However, as $Wi$ decreases, the stress contributions become increasingly dominated by the small number of highly elongated rings.
This is fundamentally different from linear melts and unprecedented for monodisperse polymers.
While linear melt structure can be described with average conformational statistics, outliers dominate the non-linear response of rings.

MD simulations provide the microscopic details necessary to identify the physics driving the strong extension of some rings at low $Wi$.
Rings are susceptible to linking with each-other to form ``reef'' or ``lark's head'' knots, shown in Figure \ref{fig:configs}(e).
These are formed when a ring that has threaded another ring then threads back through itself, as in Figure \ref{fig:configs}(d).
We observe that these links can form when chains are still in compact globular conformations.
However, once linked, rings form supramolecules with effectively higher molecular weights and relaxation times, becoming more susceptible to elongation.

We directly observe tight-links in steady-state simulations by applying primitive path analysis (PPA) to our MD trajectories \cite{Everaers2004}.
Adapting techniques from prior ring studies \cite{halverson11b,Ge2016}, we take a snapshot of the system and fix the coordinates of a single bead on each ring that is closest to the ring's geometric center.
Intramolecular repulsive interactions are then removed, driving rings to collapse into points unless they are topologically linked.
Collapsed rings have radii of gyration $R_g\sim 1$, much smaller than the average $R_g^{eq}\approx 7.14$ in equilibrium.
We measure an upper bound $\phi_{top}$ for the fraction of topologically linked rings by eliminating any ring that collapses to $R_g<0.5R_g^{eq}<3.57$, which is significantly smaller than the smallest ring conformations observed in equilibrium.
While not a direct count of link numbers, this protocol identifies the fraction of chains participating in elongated clusters that are mediated by links (Fig. \ref{fig:configs}(a)\&(b)) and that contribute significantly to the stress.

Figure \ref{fig:configs}(a) shows several highly elongated clusters of rings from the terminal state of the lowest $Wi=0.16$ flow.
Figure \ref{fig:configs}(b) shows the same rings after PPA, revealing many tight-links connecting rings into supramolecules.
Figure \ref{fig:configs}(c) shows a close up of a tight-link between two rings before (left) and after (right) PPA. Notably, supramolecules are not limited to single links, with Figure \ref{fig:configs}(a) and (b) showing supramolecules of 6, 5, and 3 rings.
In fact, we expect nonlinear response to be dominated by larger supramolecules at $Wi=0.16$, since they must be be large enough to be elongated by the low strain rate.
We suspect that the long delay in viscous thickening at low $Wi$ is due to a long induction time for larger supramolecules to spontaneously form and elongate.

The effective relaxation times of supramolecules in flow should be significantly larger than for individual rings.
Every linked chain adds $N$ monomers to a supramolecule, and we would expect the effective Rouse time of $q$ linked molecules to scale as $\sim q^2$.
Thus, linked rings experience a much larger drag in flow than unlinked rings at the same $\dot{\varepsilon}$, leading to their strong extension.
Once elongated, links are pulled tight into reef knots (Figure \ref{fig:configs}(c)), stabilizing them for as long as flow persists.
The ring topology is also important for stabilizing links. 
Unlike the recently observed transient hooking of rings by linear chains \cite{Zhou2019}, ring-ring links cannot be destroyed by convective-constraint release because rings lack ends.

\begin{figure}[ht]
    \centering
    \includegraphics[width=\linewidth]{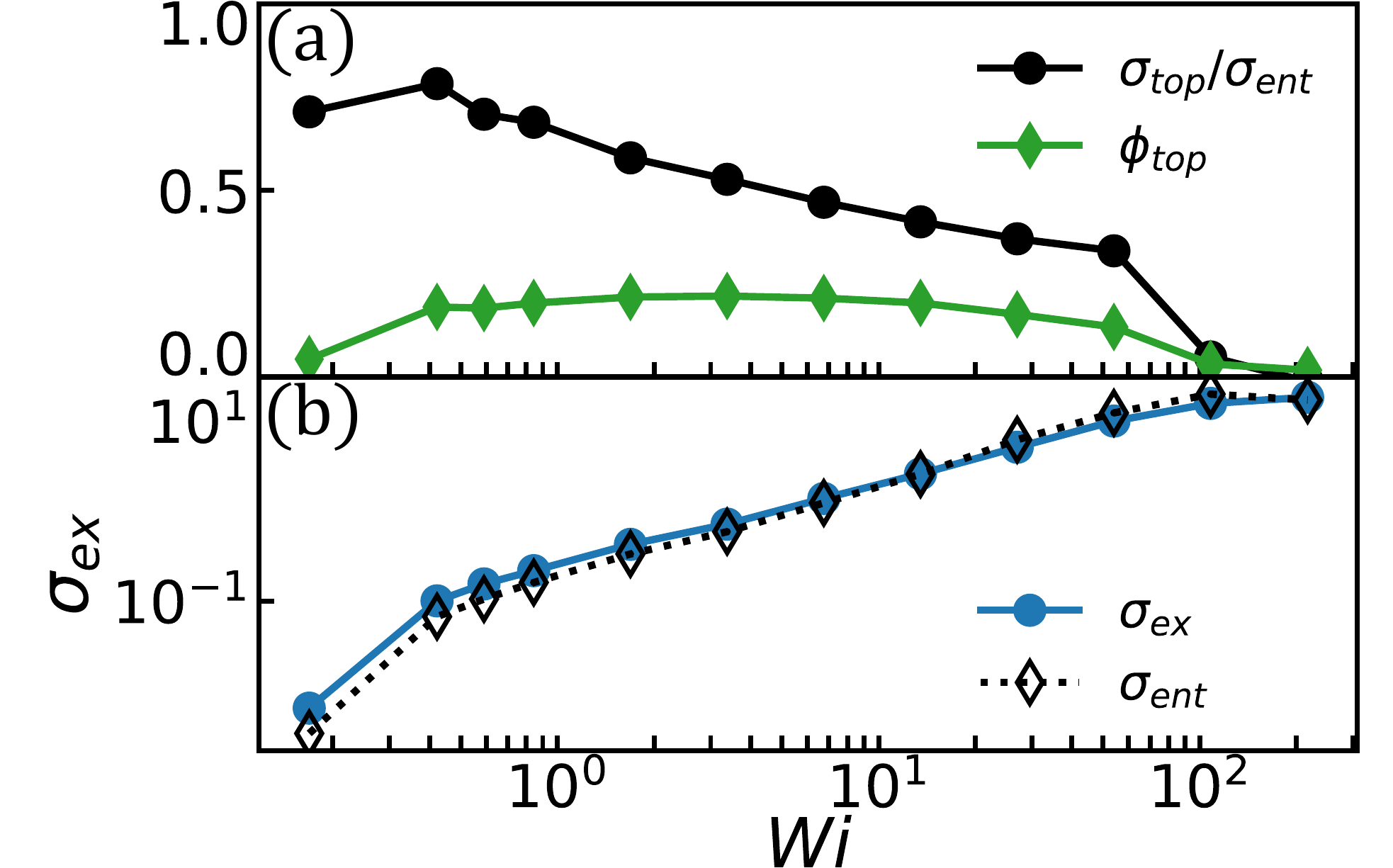}
    \caption{
    (a) Fraction of topologically linked rings $\phi_{top}$ and fraction of the entropic stress $\sigma_{top}/\sigma_{ent}$ they contribute versus $Wi$ for $N=400$ rings.
    (b) Steady-state stress $\sigma_{ex}$ versus $Wi$ compared to $\sigma_{ent}$ from Eq. \ref{eq:entropy} for an $N=400$ ring melt.
    \label{fig:ent}
    }
\end{figure}

Figure \ref{fig:ent}(a) plots the fraction of topologically linked rings $\phi_{top}$ at the terminal strain for all $Wi$ of the $N=400$ melt.
$\phi_{top}\sim0.005$ at $Wi=0.16$ and increases with increasing $Wi$ up to $\phi_{top}\sim0.20$ for $Wi>1$. This persists until $Wi>50$, where $\phi_{top}$ rapidly decreases to $<0.01$.
The sudden decrease at $Wi\approx 50$ corresponds to $\dot{\varepsilon}\tau_e\approx0.75$, with $\dot{\varepsilon}$ approaching the reciprocal of the fundamental loop relaxation time \cite{Ge2016}.  
These trends further imply the existence of an induction time for rings to sample threaded conformations that can form links.
Once $\dot{\varepsilon}\tau_e\sim1$, we expect rings elongate faster than they can sample such link-forming configurations.

Following previous studies \cite{OConnor2018PRL}, we can approximate an entropic extensional stress $\sigma_{ent}$ contributed by elongated rings:
\begin{equation}
    \sigma_{ent}=\frac{\rho k_B T}{n_k} \left< \frac{2R}{Nb} \mathcal{L}^{-1}\left( \frac{2R}{Nb}\right) P_2 \right>
    \label{eq:entropy}
\end{equation}
Here, $\rho$ is the monomer density, $n_k$ is the Kuhn segment length of the chains and $P_2$ is the nematic orientational order parameter for the ring span $\vec{R}$ relative to the extension axis.
O'Connor et al. \cite{OConnor2018PRL} found that Eq. \ref{eq:entropy} applied to entanglement segments in linear melts predicted the steady $\sigma_{ex}$ for a wide range of $Wi$.
Figure \ref{fig:ent}(b) compares $\sigma_{ent}$ to $\sigma_{ex}$ for the terminal state of the $N=400$ ring melt for all $Wi$.
The entropic prediction closely follows $\sigma_{ex}$ and agrees especially well at large $Wi$, as $P(R)$ (Figure \ref{fig:rdist}) becomes a narrow distribution.
Deviations are maximum $~30-40$\% at $Wi=0.16$; however, this $Wi$ has not reached a steady-state, so $P(R)$ and $\sigma_{ent}$ are not yet well defined.

The population $\phi_{top}$ drives the nonlinear rise in $\sigma_{ex}$ at low $Wi$.
This is seen by measuring the entropic stress contributed by just the topologically linked rings $\sigma_{top}$ by only including the $\phi_{top}$ population in the expectation value of Eq. \ref{eq:entropy}.
The fraction $\sigma_{top}/\sigma_{ent}$ is shown as circles in Figure \ref{fig:ent}(a). 
Consistent with Figure \ref{fig:rdist}, the decreasing fraction $\phi_{top}$ of linked rings contributes increasingly to the stress as $Wi$ becomes small.
Indeed, at $Wi=0.16$, linked chains comprising less than $0.5$\% of the system contribute $\sim 75$\% of $\sigma_{ent}$.
In other words, supramolecules contribute most of the anomalous stress observed at low $Wi$.
$\sigma_{top}/\sigma_{ent}$ decreases with increasing $Wi$ as $\phi_{top}$ saturates and then rapidly decreases.
Unlinked rings also elongate and contribute substantially to the stress at as $Wi$ increases, further reducing $\sigma_{top}/\sigma_{ent}$.

To conclude, MD simulations reproduce the anomalous sensitivity of ring melts to extensional flows at low $Wi$.
Anomalous increases of $\eta_{ex}$ at low $Wi$ are caused by a fraction $\phi_{top}$ of rings that PPA reveals are topological linking to form supramolecular chains that elongate at lower $Wi$ than single rings.
Microscopic measures of ring entropy directly relate the nonlinear rise in stress to elongated supramolecular clusters.  
Unlike the transient ``hooking'' of rings by linear chains \cite{Zhou2019}, these new ring-ring links are pulled tight and stabilized by the flow.
Thus, supramolecules can persist even if $Wi$ is changed.
Our future efforts will study the dynamics of topological link formation during transient flow and stress relaxation.
This will require adapting new ring analysis methods \cite{Smrek2019,Schram2019} to accurately identify and track links as they form.

\begin{acknowledgments}
M. R. acknowledges financial support from National Science Foundation under Grant No.
EFMA-1830957, the National Institutes of Health under Grant Nos. P01-HL108808, R01-HL136961, and 5UH3HL123645, and the Cystic Fibrosis Foundation. This work was supported by the Sandia Laboratory Directed
Research and Development Program.
This work was performed, in part, at the Center for Integrated Nanotechnologies, an Office of Science User Facility operated for the U.S. Department of Energy (DOE) Office of Science. Sandia National Laboratories is a multi-mission laboratory managed and operated by National Technology \& Engineering Solutions of Sandia, LLC, a wholly owned subsidiary of Honeywell International, Inc., for the U.S. DOE’s National Nuclear Security Administration under contract DE-NA-0003525. The views expressed in the article do not necessarily represent the views of the U.S. DOE or the United States Government.
\end{acknowledgments}
%\bibliography{references}% Produces the bibliography via BibTeX.
%

\end{document}